\documentstyle[epsfig]{mn2e}
\input{epsf}

\onecolumn
 
\renewcommand{\d}{{\rm d}}
\newcommand{\beq}{\begin{equation}}
\newcommand{\eeq}{\end{equation}}
\newcommand{\beqa}{\begin{eqnarray}} 
\newcommand{\eeqa}{\end{eqnarray}}
\newcommand{\bea}{\begin{array}} 
\newcommand{\ea}{\end{array}} 
\newcommand{\lag}{\langle}
\newcommand{\rag}{\rangle}
\newcommand{\Om}{\Omega_{\rm m}}
\newcommand{\Ol}{\Omega_{\Lambda}}
\newcommand{\De}{{\cal D}}

\newcommand{\Map}{M_{\rm ap}}
\newcommand{\bx}{{\bf x}}
\newcommand{\bk}{{\bf k}}
\newcommand{\kpar}{k_{\parallel}}
\newcommand{\kperp}{\bk_{\perp}}
\newcommand{\kperpDt}{k_{\perp}\De\theta_s}

\newcommand{\wt}{\tilde{w}}

\newcommand{\gammais}{\gamma_{i{\rm s}}}
\newcommand{\Mapone}{M_{\rm ap1}}
\newcommand{\Maptwo}{M_{\rm ap2}}
\newcommand{\Mapthree}{M_{\rm ap3}}

\newcommand{\cM}{{\cal M}}
\newcommand{\cH}{{\cal H}}

\newcommand{\cR}{{\cal R}}

\def\e{\epsilon}

\title[Covariance of Weak Lensing Observables]
{Covariance of Weak Lensing Observables}
\author[Munshi \& Valageas]
{Dipak Munshi$^{1,2}$, Patrick Valageas$^{3}$\\
$^{1}$Institute of Astronomy, Madingley Road,
Cambridge, CB3 OHA, United Kingdom\\
$^{2}$Astrophysics Group, Cavendish Laboratory, Madingley Road, 
Cambridge CB3 OHE, United Kingdom\\
$^{3}$ Service de Physique Th\'eorique, 
CEA Saclay, 91191 Gif-sur-Yvette, France\\}

\begin{document}
\maketitle

\begin{abstract}
Analytical expressions for covariances of weak lensing statistics related to 
the aperture mass $\Map$ are derived for 
realistic survey geometries such as SNAP for a range of 
smoothing angles and redshift bins. We incorporate the contributions to the
noise due to the intrinsic ellipticity distribution and the
effects of finite size of the catalogue. Extending previous results 
to the most general case where the overlap of source populations is
included in a complete analysis of error estimates, we study
how various angular scales in various redshifts are correlated and how the 
estimation scatter changes with survey parameters. Dependence on cosmological 
parameters and source redshift distributions are studied in detail. 
Numerical simulations are used to test the validity of various
ingredients to our calculations. Correlation coefficients are defined 
in a way that makes them practically independent of cosmology.
They can provide important tools to cross-correlate one or more different 
surveys, as well as various redshift bins within the same survey or various 
angular scales from same or different surveys. Dependence of these 
coefficients on various models of underlying mass correlation hierarchy is 
also studied. Generalisations of these coefficients at the level of 
three-point statistics have the potential to probe the complete shape 
dependence of the underlying bi-spectrum of the matter distribution. A 
complete error analysis incorporating all sources of errors suggest 
encouraging results for studies using future space based weak lensing surveys 
such as SNAP.
\end{abstract}


\section{Introduction}

Detection of weak lensing signals by observational teams 
(e.g., Bacon, Refregier \& Ellis 2000, Hoekstra et al. 2002,
Van Waerbeke et al. 2000, and Van Waerbeke et al. 2002) has led to a new 
avenue not only in constraining the background dynamics of the universe but 
in probing the nature of dark matter and dark energy as well. To do so one
compares observations with theoretical results obtained from simulations
or analytical methods.


Often numerical simulations are employed which use
ray-tracing techniques as well as line of sight integration of cosmic shear
(e.g., Schneider \& Weiss, 1988, Jarosszn'ski et al., 1990, 
Wambsganns, Cen \& Ostriker, 1998, Van Waerbeke, Bernardeau \& Mellier, 1999, 
and Jain, Seljak \& White, 2000, Couchman, Barber \& Thomas 1999) to
test analytical calculations of cosmic shear. Of course simulations are 
being constantly updated to match upcoming surveys with broader sky coverage.


Analytical techniques too have made progress in understanding of
shear induced by line-of-sight density inhomogeneities. Typically
cosmological perturbation theory is employed at large angular scales
(e.g., Villumsen, 1996, 
Stebbins, 1996, Bernardeau et al., 1997, Jain \& Seljak, 1997, Kaiser, 1998, 
Van Waerbeke, Bernardeau \& Mellier, 1999, and Schneider et al., 1998) while 
techniques based on the hierarchical {\em ansatz} provide a good match 
to simulation results at smaller angular scales
(e.g., Fry 1984, Schaeffer 1984, Bernardeau \& Schaeffer 1992, 
Szapudi \& Szalay 1993, 1997, Munshi, Melott \& Coles 1999, 1999a, 1999b). 
Ingredients 
for such calculations include Peacock \& Dodds (1996)'s prescription 
(see Peacock \& Smith (2000) for a more recent fit) for the evolution of the 
power spectrum or equivalently the two-point correlation function. Recent 
studies report an excellent agreement between analytical results and 
numerical simulations of weak lensing effects (Valageas 2000a \& b; 
Munshi \& Jain 2000 \& 2001; Munshi 2000; Bernardeau \& Valageas 2000; 
Valageas, Barber \& Munshi 2004; Barber, Munshi \& Valageas 2004; 
Munshi, Valageas \& Barber 2004). One the other hand, one may also use
halo models for specific purposes (e.g. Takada \& Jain 2003).


In realistic survey scenarios, weak lensing signals are accompanied
by various noises due to the intrinsic ellipticity distribution of galaxies, 
shot-noise due to the discreet nature of the source galaxies and finite 
volume effects due to finite survey size. A detailed analysis of these
sources of noises was carried out by Munshi \& Coles (2003) following 
Schneider et al. (1998). Recent studies by 
Valageas, Munshi \& Barber (2004) have used such techniques to model errors in
future surveys such as SNAP. It was also possible to study cross-correlation
among various surveys using such a technique. Recent analysis by 
Munshi \& Valageas (2004) showed that cross-correlation studies 
between two different surveys will have the potential to 
detect systematics associated with weak lensing measurements. 
In Munshi \& Valageas (2004) we concentrated in dividing 
the source population into two non-overlapping subsamples. 
Generalising these studies here we are also able to probe 
estimation error associated with more than two different 
smoothing angular scales for same or different source redshift distributions.
Besides the generalisation of our studies to three-point statistics means
that we are able to probe the whole bi-spectrum and to use it to
constrain theories of primordial non-Gaussianity or non-Gaussianity induced
by gravitational clustering. 

This paper is organised as follows: in section~2, a very brief 
introduction to our notations is provided. The window functions
used in the rest of the paper are introduced along with a simple model of
correlations hierarchy. In section~3, we 
take into account these realistic sources of noise to study the scatter
associated with various estimators. Specific survey 
characteristics based on SNAP class experiments are assumed for such an
analysis. In section~4 we check the validity of our results against
numerical simulations for a range of redshift distributions and
smoothing angular scales. We also investigate the dependence of two-point
and three-point correlations on large-scale structures. Finally section~5 is 
left for a discussion of our results. Appendix A provides details of 
the analytical results used in the main sections.

\section{Notations and Formalism}
\label{notations}

Any weak lensing effect $X$ smoothed over some angular radius $\theta_s$ and
averaged over the redshift distribution of sources $n(z_s)$ can be written in
terms of the fluctuations of the density field as:
\beq
X= \int\d{\vec \vartheta}\; U_X({\vec \vartheta}) \; \int_0^{\chi_{\rm max}} 
\d\chi \; \wt(\chi) \; \delta(\chi,\De{\vec \vartheta}) , \;\;\; \mbox{with} 
\; \; \; \wt(\chi) = \frac{3\Om}{2} \int_z^{z_{\rm max}} \d z_s \; n(z_s) \;
\frac{H_0^2}{c^2} \; \frac{\De(\chi) \De(\chi_s-\chi)}{\De(\chi_s)} \; (1+z) .
\label{X}
\eeq
Here the redshift $z$ corresponds to the radial distance $\chi$ and $\De$ is 
the angular distance, ${\vec \vartheta}$ is the angular direction on the sky, 
$\delta(\chi,\De{\vec \vartheta})$ is the matter density contrast and 
hereafter we normalize the mean redshift distribution of the sources 
(e.g. galaxies) to unity: $\int\d z_s \; n(z_s)=1$. 
We note $z_{\rm max}$ the depth of the survey
(i.e. $n(z_s)=0$ for $z_s>z_{\rm max}$). Here and in the following we use the 
Born approximation which is well-suited to weak-lensing studies: the 
fluctuations of the gravitational potential are computed along the 
unperturbed trajectory of the photon (Kaiser 1992). We also neglect the 
discrete effects due to the finite number of galaxies. They can be obtained 
by taking into account the discrete nature of the distribution $n(z_s)$. 
This gives corrections of order $1/N$ to higher-order moments of weak-lensing 
observables, where $N$ is the number of galaxies within the circular field 
of interest. In practice $N$ is much larger than unity (for a circular 
window of radius 1 arcmin we expect $N \ga 100$ for the SNAP mission) 
therefore in this paper we shall work with eq.(\ref{X}). The angular
filter $U_X$ depends on the weak lensing observable one considers. For 
instance, the filter associated with the aperture-mass $\Map$ is 
(Schneider 1996):
\beq
U_{\Map} = \frac{\Theta(\vartheta<\theta_s)}{\pi\theta_s^2}
\; 9 \left(1-\frac{\vartheta^2}{\theta_s^2}\right) 
\left(\frac{1}{3} - \frac{\vartheta^2}{\theta_s^2}\right) ,
\label{UX}
\eeq
where $\Theta$ is a top-hat with obvious notations. The angular radius 
$\theta_s$ gives the angular scale probed by these smoothed observables.
As described in Munshi et al. (2004), the cumulants of $X$ can be
written in real space as:
\beq
\lag X^p\rag_c = \int_0^{\chi_{\rm max}} \prod_{i=1}^{p} \d\chi_i \; 
\wt(\chi_i) \int \prod_{j=1}^{p} \d{\vec \vartheta}_j \; 
U_X({\vec \vartheta}_j) \;\; 
\xi_p\left( \bea{l} \chi_1 \\ \De_1 {\vec \vartheta}_1 \ea ,
\bea{l} \chi_2 \\ \De_2 {\vec \vartheta}_2 \ea , \dots , 
\bea{l} \chi_p \\ \De_p {\vec \vartheta}_p \ea \right) ,
\label{cumXr}
\eeq
or equivalently we can write in Fourier space:
\beq
\lag X^p \rag_c = \int_0^{\chi_{\rm max}} \prod_{i=1}^{p} \d\chi_i \; 
\wt(\chi_i) \int \prod_{j=1}^{p} \d\bk_j \; W_X(\bk_{\perp j} \De_j \theta_s) 
\;\; \left( \prod_{l=1}^{p} e^{i k_{\parallel l} \chi_l} \right) 
\;\; \lag \delta(\bk_1) \dots \delta(\bk_p) \rag_c .
\label{cumXk}
\eeq
We note $\lag .. \rag$ the average over different realizations of the
density field, $\xi_p$ is the real-space $p-$point correlation function of 
the density field $\xi_p(\bx_1,..,\bx_p)= \lag \delta(\bx_1) .. 
\delta(\bx_p)\rag_c$, $\kpar$ is the component of $\bk$ parallel to the 
line-of-sight, $\kperp$ is the two-dimensional vector formed by the 
components of $\bk$ perpendicular to the line-of-sight and 
$W_X(\kperp\De\theta_s)$ is the Fourier transform of the window $U_X$:
\beq
W_X(\kperp\De\theta_s) = \int\d{\vec \vartheta} \; U_X({\vec \vartheta}) 
\; e^{i \kperp.\De{\vec \vartheta}} , \;\;\; \mbox{whence} \;\;\; 
W_{\Map}(\kperp\De\theta_s) =  \frac{24 J_4(\kperpDt)}{(\kperpDt)^2} .
\label{WX}
\eeq
The real-space expression (\ref{cumXr}) is well-suited
to models which give an analytic expression for the correlations $\xi_p$,
like the minimal tree-model (Valageas 2000b; Bernardeau \& Valageas 2002; 
Barber et al. 2004) while the Fourier-space expression (\ref{cumXk}) is 
convenient for models which give a simple expression for the correlations
$\lag \delta(\bk_1) .. \delta(\bk_p)\rag_c$, like the stellar model 
(Valageas et al. 2004; Barber et al. 2004). In this article we shall use
the latter stellar model defined by:
\beq
\lag \delta(\bk_1) .. \delta(\bk_p)\rag_c = \frac{\tilde{S}_p}{p} \; 
\delta_D(\bk_1+\dots+\bk_p) \; \sum_{i=1}^p \prod_{j\neq i} P(k_j) , \;\;\;
\mbox{with} \;\;\; \tilde{S}_2=1 ,
\label{stellar}
\eeq
where $\delta_D$ is the Dirac distribution and $P(k)$ is the 3-d 
power-spectrum of the density fluctuations. The coefficients $\tilde{S}_3,
\tilde{S}_4,..$ are closely related (and approximately equal) to the skewness,
kurtosis, .., of the density field. Then, substituting 
eq.(\ref{stellar}) into eq.(\ref{cumXk}) yields (Munshi et al. 2004):
\beq
\lag X^p \rag_c = \tilde{S}_p \int_0^{\chi_{\rm max}} \d\chi \; \wt^p 
\int\d{\vec \vartheta} \; U_X({\vec \vartheta}) \;
I_X(\chi,{\vec \vartheta})^{p-1} , \;\;\; \mbox {with} \;\;\;
I_X(\chi,{\vec \vartheta}) = \frac{1}{2} \int \frac{\d\kperp}{k_{\perp}^2} \;
\frac{\Delta^2(k_{\perp},z)}{k_{\perp}} \; W_X(\bk_{\perp j} \De_j \theta_s)
\; e^{-i \kperp.\De{\vec \vartheta}} .
\label{cumstellar}
\eeq
Here we introduced the power per logarithmic wavenumber 
$\Delta^2(k)=4\pi k^3P(k)$. Of course, these expressions generalize in a
straightforward fashion for many-point cumulants. For instance, for two-point
cumulants we obtain within this model:
\beq
\lag X_1^p X_2^q \rag_c = \tilde{S}_{p+q} \int_0^{\chi_{\rm max}} \d\chi \; 
\wt_1^p \wt_2^q \int\d{\vec \vartheta} \; \left\{ \frac{p}{p+q} \; 
U_1({\vec \vartheta}) \; I_1(\chi,{\vec \vartheta})^{p-1} \;
I_2(\chi,{\vec \vartheta})^q + \frac{q}{p+q} \; U_2({\vec \vartheta}) \; 
I_1(\chi,{\vec \vartheta})^p \; I_2(\chi,{\vec \vartheta})^{q-1} \right\} .
\label{cum2stellar}
\eeq
These quantities describe the cross-correlations between two surveys or 
subsamples, which are refered to by the subscripts ``i=1,2''. The source
redshift distributions $n_i(z_s)$ and the smoothing angles $\theta_{si}$
can be different, as well as the directions on the sky. However, in this
paper we shall restrict ourselves to correlations between angular cells which
are centered on the same direction. Then there is no need to add an angular 
shift to the expressions (\ref{UX}) and (\ref{WX}) of the filter functions.
On the other hand, note that the signal would decrease for nonzero angular 
separations. In this article we shall consider both two-point and three-point 
cumulants. The latter contain valuable information about the bispectrum and 
can be used as a template for non-Gaussianity studies.

The weak lensing effects associated with different redshift bins or angular
scales are correlated since their lines of sight probe the same density
fluctuations at low $z$ (where they are largest). In order to measure these 
cross-correlations we define the cross-correlation coefficients $r_{pq}$ as:
\beq
r_{pq} = \frac{\lag X_1^p X_2^q \rag_c}{\lag X_1^{p+q}\rag_c^{p/(p+q)}
\lag X_2^{p+q}\rag_c^{q/(p+q)}} \; , \;\;\; \mbox{in particular}
\;\;\; r_{11} = \frac{\lag X_1 X_2 \rag_c}{\lag X_1^2\rag_c^{1/2}
\lag X_2^2\rag_c^{1/2}}  .
\label{rpq}
\eeq
The quantities $r_{pq}$ correspond to the two-point cumulants 
$\lag X_1^p X_2^q \rag_c$ normalized in such a way that most of the dependence
on cosmology and gravitational dynamics cancels out. Thus, if the two 
subsamples are highly correlated we have $r_{pq} \simeq 1$ while 
$r_{pq} \simeq 0$ if they are almost uncorrelated. In a similar fashion we 
can generalize $r_{pq}$ to three-point objects $r_{pqs}$ as:
\beq
r_{pqs} = \frac{\lag X_1^p X_2^q X_3^s \rag_c}
{\lag X_1^{p+q+s}\rag_c^{p/(p+q+s)} \lag X_2^{p+q+s}\rag_c^{q/(p+q+s)} 
\lag X_3^{p+q+s}\rag_c^{s/(p+q+s)}} \; , \;\;\; \mbox{in particular}
\;\;\; r_{111} = \frac{\lag X_1 X_2 X_3 \rag_c}{\lag X_1^3\rag_c^{1/3}
\lag X_2^3\rag_c^{1/3} \lag X_3^3\rag_c^{1/3}} .
\label{rpqs}
\eeq
These objects provide valuable information about the bi-spectrum as they
directly probe three different angular scales over three different redshift
distributions. They can also be used to study cross-correlations among three 
different surveys some of which may have overlapping source distributions. 
However note that as a result of subdividing the same source samples into 
various redshift bins the shot-noise will increase. In addition,
being a true three-point object it will be more affected by the finite size of
the survey volume at larger angular scales. However a good sky coverage
for future surveys such as SNAP will mean these objects can be measured 
over reasonable angular length scales. It is possible to investigate even
higher numbers of points (samples) but we shall not consider this here.

\section{Future Surveys}
\label{Future Surveys}

\subsection{A specific case study: Wide SNAP survey}
\label{Wide SNAP}

In order to illustrate the measurements one can expect from future surveys,
we consider here the case of the Wide SNAP survey. We use the characteristics 
of the SNAP mission as given in Refregier et al. (2004). The redshift 
distribution of galaxies and the total surface density $n_g$ of usable 
galaxies are:
\beq
n(z_s) \propto z_s^2 \; e^{-(z_s/z_0)^2} \;\; \mbox{with} \;\; z_0=1.13 , 
\;\; z_{\rm max}=3 \;\;\; \mbox{and} \;\;\; n_g= 100 \; \mbox{arcmin}^{-2} .
\label{nzSNAP}
\eeq
The shear variance due to intrinsic ellipticities and measurement errors is 
$\sigma_*=\lag|\epsilon_*|^2\rag^{1/2}=0.31$ while the survey covers an area 
$A=300\;\mbox{deg}^2$. Therefore, we take for the number $N$ of galaxies within
a circular field of radius $\theta_s$ and for the number $N_c$ of cells of 
radius $\theta_s$:
\beq
N= n_g\pi\theta_s^2 \simeq 314 \left( \frac{n_g}{100 \mbox{arcmin}^{-2}} 
\right)  \left(\frac{\theta_s}{1\mbox{arcmin}}\right)^2 \;\;\; \mbox{and} 
\;\;\; N_c= \frac{A}{(2\theta_s)^2} = 2.7 \times 10^5 
\left(\frac{A}{300\mbox{deg}^2}\right) 
\left(\frac{\theta_s}{1\mbox{arcmin}}\right)^{-2} .
\label{Nsurvey}
\eeq
For some purposes we shall also consider the two subsamples which can be 
obtained from the Wide SNAP survey by dividing galaxies into two redshift 
bins: $z_s>z_*$ (which we refer to as ``Wide$>$'') and $z<z_*$ (``Wide$<$''). 
We choose $z_*=1.23$, which corresponds roughly to the separation provided by 
the SNAP filters and which splits the Wide SNAP survey into two samples 
with the same number of galaxies (hence $n_g=50$ arcmin$^{-2}$).

For the background cosmology we consider a LCDM model with $\Om=0.3$, 
$\Ol=0.7$, $H_0=70$ km/s/Mpc and $\sigma_8=0.88$.

\subsection{Two-point correlations for $\Map$}
\label{Two-point correlations for Map}

\begin{figure}
\protect\centerline{
\epsfysize = 2.3 truein
\epsfbox[22 519 591 705]
{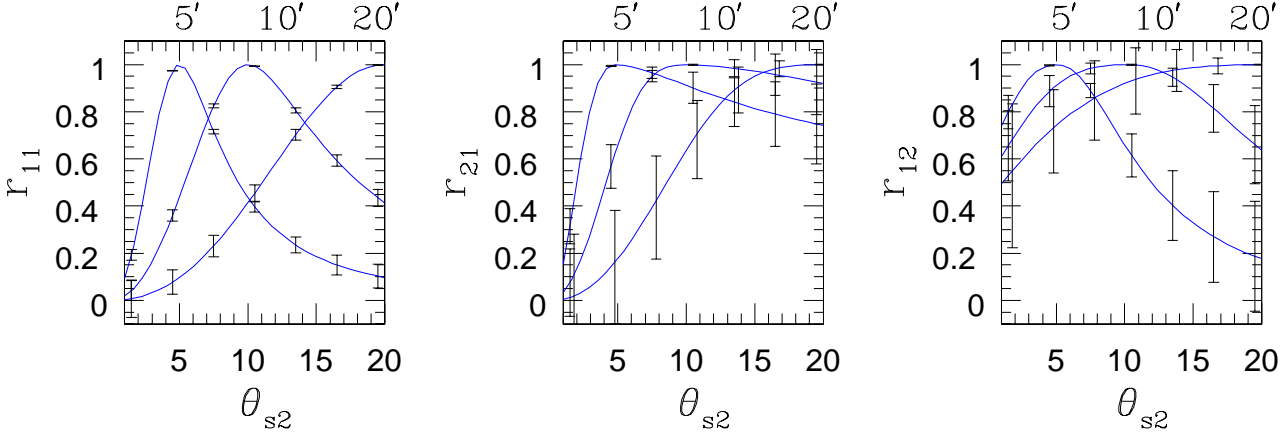}}
\caption{Correlation coefficients $r_{11}$ (left panel), $r_{21}$ 
(middle panel) and $r_{12}$ (right panel) are plotted
as a function of the smoothing angle $\theta_{s2}$. All cases
correspond to the full SNAP survey. The three different curves correspond to
the smoothing angle $\theta_{s1}=5',10'$ and $20'$, whereas $\theta_{s2}$
runs from $1'$ to $20'$. Errorbars associated with each curve
show the 1-$\sigma$ scatter in the measured values, due to intrinsic 
ellipticities and cosmic variance.}
\label{figrpq_snap}
\end{figure}

\begin{figure}
\protect\centerline{
\epsfysize = 2.3 truein
\epsfbox[22 519 591 705]
{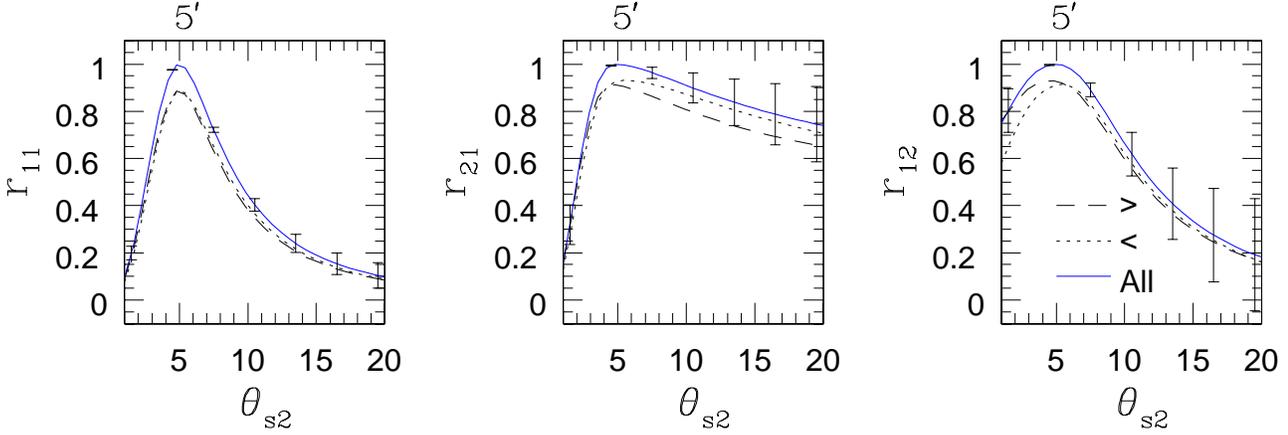}}
\caption{Correlation coefficients $r_{11}$, $r_{21}$ and $r_{12}$ are plotted 
for the wide SNAP survey with redshift binning. Solid curves denote the 
complete source sample whereas other line-styles correspond to redshift 
binning. The dashed curve ``$>$'' shows the case where $\Mapone$ at fixed 
$\theta_{s1}$ is measured from high-$z$ galaxies ($z_{s1}>1.23$) while
$\Maptwo$ at $\theta_{s2}$ which runs from $1'$ to $20'$ is measured from
low-$z$ galaxies ($z_{s2}<1.23$). The dotted curve ``$<$'' corresponds to
$z_{s1}<1.23$ and $z_{s2}>1.23$.}
\label{figrpq_snap_redbin}
\end{figure}

In this article we shall restrict ourselves to the statistics obtained
from the aperture-mass $\Map$, defined from the filters $U_{\Map}$ and
$W_{\Map}$ given in eqs.(\ref{UX}), (\ref{WX}). We first show in  
Fig.~\ref{figrpq_snap} the lowest-order two-point correlation coefficients 
$r_{pq}$ obtained at a fixed smoothing angle $\theta_{s1}$ while 
$\theta_{s2}$ runs from $1'$ up to $20'$, for the full wide SNAP survey. 
The three curves in each panel correspond to $\theta_{s1}=5',10'$ and $20'$. 
The error-bars take into account the cosmic variance and the intrinsic 
ellipticities of galaxies, see Appendix and Munshi \& Valageas (2004).

The correlation coefficients $r_{pq}$ show a peak where both scales coincide
and are fully correlated ($r_{pq}=1$ as $\Mapone$ and $\Maptwo$ are actually
identical at this point). We find from $r_{11}$ that the correlation remains 
strong 
($\ga 0.4$) up to an angular scale $\theta_{s2}$ twice larger than the other 
smoothing scale $\theta_{s1}$. We must point out that this depends on the 
smoothing window and that top-hat windows (for the smoothed convergence 
$\kappa_s$, or the shear components $\gammais$) generate more extended 
correlation profiles. This is related to the fact that the aperture-mass
$\Map$ is a very localized filter in Fourier space (Schneider 1996). However,
we can see that the third-order correlations $r_{21}$ and $r_{12}$ decrease
much more slowly (especially $r_{21}$). 
It is interesting to note that the shapes of $r_{21}$ 
and $r_{12}$ are not identical and that $r_{21}$ exhibits in particular
a very extended tail at larger angles. This can be understood from 
eq.(\ref{cumXk}). The Dirac factor $\delta_D(\bk_1+\bk_2)$ contained in the
mean $\lag\delta(\bk_1)\delta(\bk_2)\rag_c$ involved in $r_{11}$ implies
$k_1=k_2$ so that $r_{11}$ peaks at $\theta_{s1}=\theta_{s2}$ where both
$W(k_1\De\theta_{s1})$ and $W(k_2\De\theta_{s2})$ can be simultaneously 
maximized. Then, $r_{11}$ decreases for different angular scales where one
cannot simultaneously maximize $W(k_1)$ and $W(k_2)$. On the other hand,
for the third-order correlation $r_{21}$ the Dirac factor is 
$\delta_D(\bk_1+\bk_1'+\bk_2)$ with a weight 
$W(k_1\De\theta_{s1})W(k_1'\De\theta_{s1})W(k_2\De\theta_{s2})$. Then, at
large $\theta_{s2}$ it is still possible to maximize all three weights
$W$ by choosing $k_2 \sim 1/\De\theta_{s2}$, 
$k_1 \sim k_1' \sim 1/\De\theta_{s1} \gg k_2$, with $\bk_1' \simeq -\bk_1$ and
$\bk_2=-\bk_1-\bk_1'$.
This is not possible for $r_{12}$ where the weight is now 
$W(k_1\De\theta_{s1})W(k_2\De\theta_{s2})W(k_2'\De\theta_{s2})$ and from
two small wavenumbers $k_2$ and $k_2'$ of order $1/\De\theta_{s2}$ it is
impossible to construct a wavenumber $\bk_1=-\bk_2-\bk_2'$ which is much
larger ($k_1$ is of order $1/\De\theta_{s2} \ll 1/\De\theta_{s1}$).
A similar reasoning also explains the steep decline of the correlation 
coefficient $r_{21}$ at low angular scales $\theta_{s2}$.
Thus, the correlation coefficients $r_{pq}$ directly probe the features of
the matter density field (note that the previous arguments only depend on
the statistical homogeneity of the density field).
We can see from Fig.~\ref{figrpq_snap} that the error-bars expected from the
wide SNAP survey are quite small for $r_{11}$. They are larger for the
third-order statistics $r_{21}$ and $r_{12}$ but one should still be able to
obtain a good measure of the shape of these correlations. On the other hand,
the measure of the coefficients $r_{pq}$ can provide a way to check the 
importance of noise and systematics. 

Next, we show in Fig.~\ref{figrpq_snap_redbin} the effect of the redshift 
binning of the data. We divide the galaxies into two different bins: 
$z_{s1}>1.23$ and $z_{s2}<1.23$ (dashed curve ``$>$'') or $z_{s1}<1.23$ and 
$z_{s2}>1.23$ (dotted curve ``$<$''). Of course, the behaviour of the various
correlation coefficients $r_{pq}$ remains similar to the case of only one
redshift bin displayed in Fig.~\ref{figrpq_snap}. However, at the peak
$\theta_{s1}=\theta_{s2}$ we now have $r_{pq}<1$ since even for the
same smoothing angle the quantities $\Mapone$ and $\Maptwo$ are not identical
as their lines of sight extend to different redshifts. Of course the 
error-bars are somewhat larger when we apply such a redshift binning of the
data but it is still possible to measure these coefficients to a reasonable 
accuracy up to third-order statistics $r_{21}$ and $r_{12}$. This probes
the evolution with redshift of gravitational clustering, although the latter
is best measured with one-point statistics (most of the time dependence of
the amplitude of density fluctuations and of the cosmological distances 
associated with various source redshifts is factorized out of the coefficients
$r_{pq}$).

\begin{figure}
\protect\centerline{
\epsfysize = 2.3 truein
\epsfbox[22 519 591 705]
{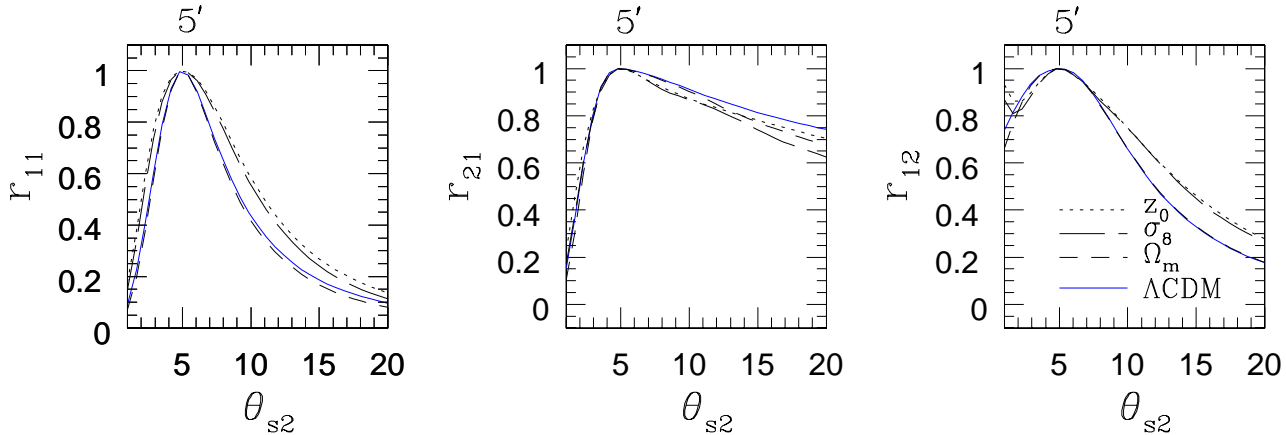}}
\caption{Correlation coefficients $r_{11}$, $r_{21}$ and $r_{12}$ are 
plotted for various cosmological or survey parameters. The solid line is our
baseline LCDM model, the ``$\Om$'' curve shows a $20\%$ increase of $\Om$
(from $\Om=0.3$ up to $\Om=0.36$), the ``$\sigma_8$'' curve is a $20\%$ 
increase of $\sigma_8$ (from $\sigma_8=0.88$ up to $\sigma_8=1.06$) and
the ``$z_0$'' curve is a $20\%$ decrease of the characteristic redshift
$z_0$ of the survey (from $z_0=1.13$ down to $z_0=0.9$), eq.(\ref{nzSNAP}).
The distance between these latter curves and the baseline model has been
{\it multiplied by a factor 20} since the dependence on cosmological or
survey parameters actually nearly cancels out.}
\label{figrpq_snap_20}
\end{figure}

Finally, we also display in Fig.~\ref{figrpq_snap_20} the results obtained
with a $20\%$ increase of $\Om$ (from $\Om=0.3$ up to $\Om=0.36$), or a $20\%$ 
increase of the normalization $\sigma_8$ of the density power-spectrum 
(from $\sigma_8=0.88$ up to $\sigma_8=1.06$), or a $20\%$ decrease of the 
characteristic redshift $z_0$ (eq.(\ref{nzSNAP})) of the survey 
(from $z_0=1.13$ down to $z_0=0.9$). The distance between these latter curves 
and the baseline model (solid line) has been {\it multiplied by a factor 
20}. Thus we see that the dependence on cosmological or survey parameters 
almost completely cancels out of the reduced correlation coefficients
$r_{pq}$. Therefore, these quantities provide a direct handle on the noise
or on the detailed properties of the large-scale density field (e.g. the
validity of a hierarchical model like (\ref{stellar}) for many-body correlation
functions). On the other hand, cosmological parameters may be measured from
one-point cumulants. Thus, the use of one-point, two-point and three-point 
cumulants provides a convenient separation between various features of the
problem: cosmology, large-scale structures and noise.

\subsection{Three-point correlations for $\Map$}
\label{Three-point correlations for Map}

\begin{figure}
\protect\centerline{ 
\epsfysize = 3. truein
\epsfbox[22 433 307 705]
{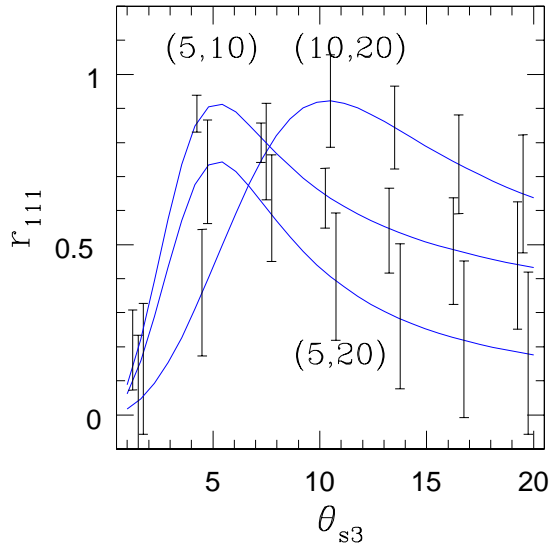}}
\caption{The correlator $r_{111}$ for the full wide SNAP survey is plotted as 
a function of smoothing angle $\theta_{s3}$ for a fixed pair of 
$(\theta_{s1},\theta_{s2})$. The pair $(\theta_{s1},\theta_{s2})$ for each 
curve is indicated in the plot. Error bars denote the $1-\sigma$ scatter 
around the mean.}
\label{figr111_snap}
\end{figure}

We show in Fig.~\ref{figr111_snap} the three-point correlation coefficient
$r_{111}$ defined in eq.(\ref{rpqs}), for the full wide SNAP survey. This 
three-point object is a direct probe of the underlying bi-spectrum of the
matter density field. The 
error-bars are computed from the expressions derived in the Appendix where 
the optimised estimators $H_{pqs}$ are used (this helps to reduce the scatter,
see Valageas et al. 2004). The statistic $r_{111}$ is a symmetric function of 
its arguments when the three source redshift distributions are the same as 
in Fig.~\ref{figr111_snap} where the full source sample is used. Then, when
two smoothing angles are equal it coincides with the two-point statistics
$r_{21}$ or $r_{12}$. As in the case of two-point correlators the error-bars 
are dominated by the finite size of the survey and are higher for larger 
smoothing angular scales. However from the figure we can see that SNAP will 
be able to probe $r_{111}$ reasonably well even when a large smoothing 
angle $\theta_{s2}=20'$ or $\theta_{s3}=20'$ is used. Redshift binning
can be useful in studying the redshift evolution of $r_{111}$ but 
this reduces the number of sources thereby increasing the shot noise which 
starts playing a dominant role at smaller smoothing angular scales.
The noise level at given smoothing angular scales is not only determined by 
the signal-to-noise ratios $\rho_1$, $\rho_2$ and $\rho_3$ of the variance
of the aperture-mass at each scale but also by the overlap 
of various smoothing patches as described by $\rho_{12}$, $\rho_{23}$ and 
$\rho_{13}$, see eq.(\ref{rho}) and the Appendix. It is possible to 
generalise our results to cases where the patches are not all centered on 
the same line of sight but we shall not investigate this point here. 
 
In  Fig.~\ref{figr111_snap} we plot $r_{111}$ for three combinations of 
$(\theta_{s1},\theta_{s2})$ while $\theta_{s3}$ runs from $1'$ to $20'$. 
The curves show a single peak reached when $\theta_{s3}$ is equal to the 
smallest angular scale among $(\theta_{s1},\theta_{s2})$. This may be
understood as follows. Let $\theta_{s1}<\theta_{s2}$, then as $\theta_{s3}$ 
increases from $0'$ the value of $r_{111}$ grows as the surface which is 
common to the three patches increases as $\theta_{s3}^2$. More to the point, 
the relative importance of the common surface increases: relative to the three
quantities $M_{{\rm ap}j}$ it goes as 
$\{(\theta_{s3}/\theta_{s1})^2,(\theta_{s3}/\theta_{s2})^2,1\}$. Next, as 
$\theta_{s3}$ gets larger than $\theta_{s1}$ the common surface sticks to 
$\theta_{s1}^2$ (it no longer grows) while the relative common surface 
actually decreases as 
$\{1,(\theta_{s1}/\theta_{s2})^2,(\theta_{s1}/\theta_{s3})^2\}$. Therefore, one
can expect a steady decline beyond $\theta_{s1}$. This is indeed what we
observe in Fig.~\ref{figr111_snap}. Of course, matters are actually a bit
more intricate as the three-point correlation is not ``$\delta_D$-correlated''
in real space so that the correlation between different patches is not
strictly proportional to their geometrical overlap and the filters $U_{\Map}$
are not simple top-hats. However, these simple arguments explain the overall
trends seen in Fig.~\ref{figr111_snap}.

As expected, for a fixed angle $\theta_{s1}$ (here $5'$) the peak of $r_{111}$
at $\theta_{s3}=\theta_{s1}$ is larger when $\theta_{s2}$ is closer to
$\theta_{s1}$ (here $10'$ as compared with $20'$). Note that $r_{111}=1$ for
$\theta_{s1}=\theta_{s2}=\theta_{s3}$ where the three quantities $\Mapone$,
$\Maptwo$ and $\Mapthree$ are identical whence fully correlated (for identical
source distributions).
We can see that the correlation of third-order statistics remains significant
over a large range of angular scales and shows a slow decline at high 
$\theta_{s3}$ and a steeper falloff at low $\theta_{s3}$, in a fashion similar
to $r_{21}$ shown in Fig.~\ref{figrpq_snap}, for the same reasons.

\section{Comparison with numerical simulations}
\label{Comparison with numerical simulations}

We have shown in the previous section that the two-point and three-point
correlators $r_{pq}$ and $r_{pqs}$ associated with the aperture-mass
$\Map$ could be measured in future surveys like SNAP and allow a clean
separation between cosmological parameters and the properties of the
density field (the behaviour of its many-body correlations). They can also
be used to monitor the noise of the survey. In this section, we address the
use of these correlation coefficients from the point of view of large-scale
structures. Thus, we first compare our predictions from the simple
hierarchical model (\ref{stellar}) with the results of numerical simulations.
Secondly, we investigate the dependence of these correlators on the detailed
properties of the density field. 

The numerical method for the computation of the lensing statistics is
based on the original formalism of Couchman, Barber \& Thomas (1999)
which concentrates on computation of three-dimensional shear matrices
along lines of sight through the linked simulation volumes. This
technique has been further developed by Barber (2002).
This procedure which has been described in various publications 
in detail (e.g. Barber 2002) has been applied to a LCDM 
cosmological simulation created by the Hydra
Consortium \footnote{http://hydra.mcmaster.ca/hydra/index.html} using
the `Hydra' $N$-body hydrodynamics code (Couchman, Thomas \& Pearce,
1995). Its cosmological parameters are: $\Om=0.3$, $\Ol=0.7$, $\Gamma=0.25$,
$H_0=93$ km/s/Mpc and $\sigma_8=1.22$. The computations of 
correlation functions were done by using concentric annuli
which are divided in equal radial bins and angular bins in real space
directly from the rectangular grid. Various combinations of bin width in radial 
and angular directions were considered while computing the correlations to 
check the stability of our scheme. Several levels of dilution were tested 
by increasing the grid size of the parent rectangular. For simplicity, we 
consider Dirac delta distributions of sources. That is, for three-point 
correlators all sources are located at the three redshifts $z_{s1},z_{s2}$ 
and $z_{s3}$ which can be different.

\subsection{Two-point correlators}
\label{Two-point correlators}

\begin{figure}
\protect\centerline{
\epsfysize = 3.0 truein
\epsfbox[22 297 591 545]
{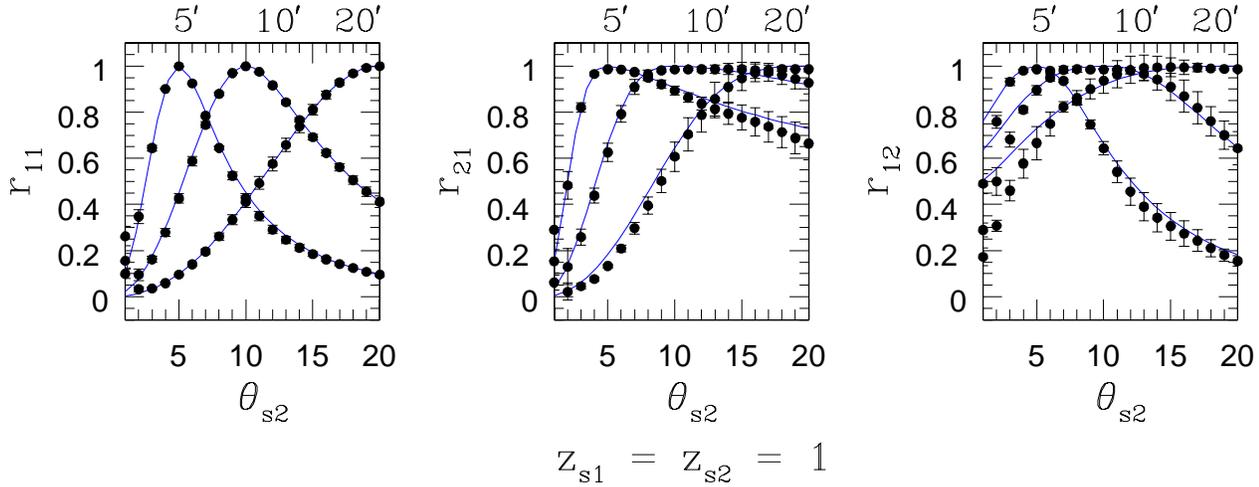}}
\caption{The two-point correlation coefficients $r_{11},r_{21}$ and $r_{12}$ 
are plotted for various angular scales. In each plot the curves from left to 
right correspond to $\theta_{s1}=5',10'$ and $20'$, while $\theta_{s2}$ runs 
from $1'$ to $20'$. Source redshift is fixed at $z_{s1}=z_{s2}=1$. 
Solid points are measurements from simulation data while the solid curve is
the theoretical prediction from the simple stellar model (\ref{stellar}).
Error-bars are computed from $10$ identical realizations. Note that the 
LCDM cosmology used for the simulations is slightly different from the one
used for the SNAP survey.}
\label{figrpq_sim}
\end{figure}

\begin{figure}
\protect\centerline{
\epsfysize = 3.0 truein
\epsfbox[22 297 591 545]
{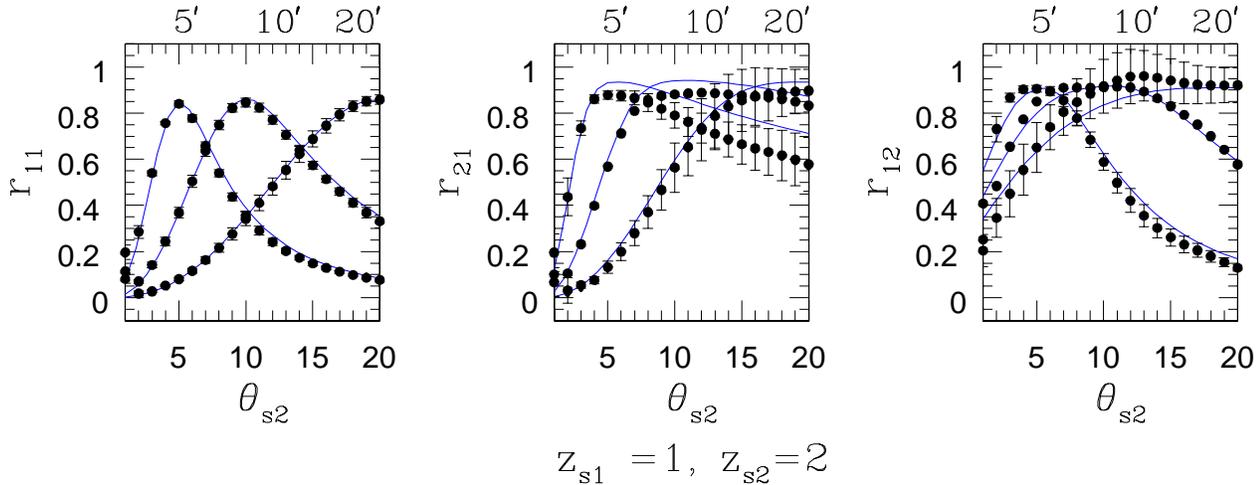}}
\caption{The two-point correlation coefficients $r_{11},r_{21}$ and $r_{12}$ 
are plotted for the angular scale $\theta_{s1} =5',10'$ and $20'$ while
$\theta_{s2}$ runs from $1'$ to $20'$. The redshift slices considered are 
$z_{s1}=1$ and $z_{s2}=2$. The solid curve is the analytical model 
(\ref{stellar}) while solid points are measurements from simulation data.}
\label{figrpq_sim_z}
\end{figure}

We again study the two-point objects such as $r_{pq}$ by keeping one angular 
scale fixed, say $\theta_{s1}$, while the other smoothing angle $\theta_{s2}$
varies from $1'$ to $20'$ (unless $p=q$ and $z_{s1}=z_{s2}$, $r_{pq}$ is not 
a symmetric function of its arguments $\theta_{s1}$ and $\theta_{s2}$). 
This range is determined by the finite resolution of the simulations: smaller
scales are affected by the rectangular grid (i.e. resolution) while large 
angular scales are affected by the finite simulation box. Note that
effects due to the finite size of the box will be more prominent at smaller 
redshifts while effects due to the grid play a more important role at larger 
redshifts. In most of our studies we analyse cross-correlations and 
auto-correlations of various redshift slices against the slice $z_s=1$ which 
is least affected by such artifacts and of great observational interest.

We display the case of identical source redshifts $z_{s1}=z_{s2}=1$ in
Fig.~\ref{figrpq_sim} and the case of different source redshifts in
Fig.~\ref{figrpq_sim_z}, with $z_{s1}=1$ and $z_{s2}=2$.
As in Figs.~\ref{figrpq_snap},~\ref{figrpq_snap_redbin}, the two-point 
correlators always show a peak where both angular scales coincide and complete 
overlap is achieved to generate a high correlation coefficient $r_{pq}$. 
Again, in cases where the same redshift slices are used for the two smoothing
angles the coefficients $r_{pq}$ reach a value of unity, whereas for different 
redshift slices the maximum is less than unity as only partial overlap of 
the lines 
of sight is achieved. We also recover the slow decrease at large angles
$\theta_{s2}$ of $r_{21}$. We can see that the simple hierarchical model
(\ref{stellar}) agrees reasonably well with the numerical simulations, 
although there is a small discrepancy at large angles $\theta_{s2}$ for 
$r_{21}$. However, the error-bars in this domain are large.
In agreement with the nearly total independence of these correlation
coefficients onto cosmology, described in Fig.~\ref{figrpq_snap_20},
we have checked that an OCDM simulation (again from the Hydra Consortium)
yields the same results.
The error-bars shown are computed from a large number of realizations 
$n_{{\rm real}}=10$. As expected they are larger for higher-order statistics
($r_{21}$ and $r_{12}$ as compared with $r_{11}$). However, note that all 
realizations we have used are actually constructed from the same N-body 
density field and are not completely independent so there could be a 
systematic deviation (although expected to be small) in our results which can 
only be studied by using completely different simulations where density fields 
probed by various lines of sights are completely independent.
In this article we only consider angular windows with zero separation. Then
the scatter is mostly determined by the largest of the smoothing angles which 
defines how many independent patches there are for correlation studies.

The effects of small box size not only appear as an increased scatter for
the estimation of high-order correlation functions it can also bias the mean.
Detailed schemes exist in case of galaxy clustering to correct or estimate
such bias from galaxy surveys or simulated catalogs.
For large smoothing angles with $\theta_{s2}=20'$ it is expected that
such effects will be visible in the computed values of $r_{12}$ estimated 
from our simulated maps as there are only few independent patches within the 
survey and $r_{12}$ puts more weight on $\theta_{s2}$ as compared with 
$r_{21}$. Clearly such effects will be more pronounced at larger $\theta_{s1}$ 
and at lower redshifts where gravitational clustering has effectively 
generated correlated patches on the sky. We have made no attempt to correct for
such a deviation. However, note that the compensated filter
associated with $\Map$, which is very localized in Fourier space, makes these
problems less acute than for simple tophat filters.

\subsection{Three-point correlators}
\label{Three-point correlators}

\begin{figure}
\protect\centerline{
\epsfysize = 2.85 truein
\epsfbox[22 433 304 715]
{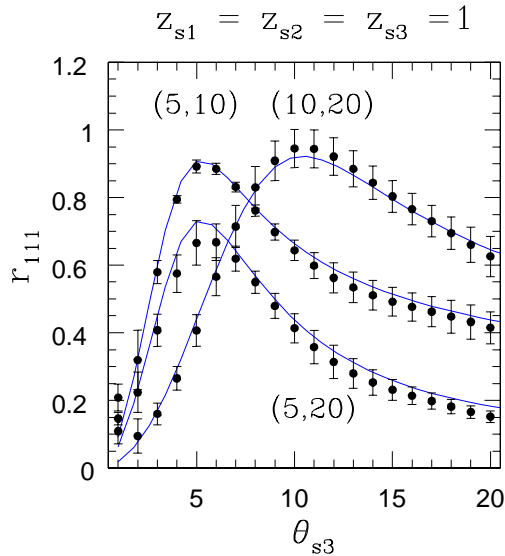}}
\caption{The three-point correlation coefficient $r_{111}$ as a function of 
smoothing angle $\theta_{s3}$ for a fixed pair of $(\theta_{s1},\theta_{s2})$.
The pair $(\theta_{s1},\theta_{s2})$ for each curve is indicated in the plot.
The three source redshifts are equal: $z_{s1}=z_{s2}=z_{s3}=1$. The solid 
curve is the analytical model (\ref{stellar}) while solid points with 
error-bars are measurements from simulation data.}
\label{figr111_sim}
\end{figure}

\begin{figure}
\protect\centerline{
\epsfysize = 4.5 truein
\epsfbox[22 127 394 555]
{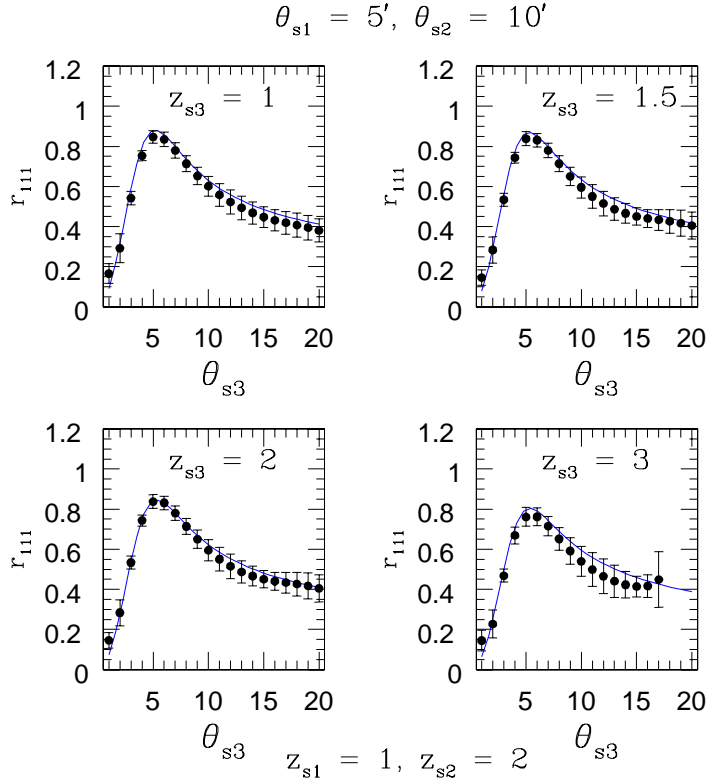}}
\caption{The three-point correlator $r_{111}$ is plotted as a function of 
smoothing angle $\theta_{s3}$ for the fixed pair 
$(\theta_{s1},\theta_{s2})=(5',10')$. The source redshifts are $z_{s1}=1$ and
$z_{s2}=2$ while we have $z_{s3}=1$ (upper left panel), $z_{s3}=1.5$ (upper 
right panel), $z_{s3}=2$ (lower left panel) and $z_{s3}=3$ (lower right panel).
The solid curve is the analytical model (\ref{stellar}) while solid points 
with error-bars are measurements from simulation data.}
\label{figr111_red_sim}
\end{figure}

We show in Fig.~\ref{figr111_sim} the three-point correlation coefficient
$r_{111}$ as a function of smoothing angle $\theta_{s3}$ for three fixed
pairs of $(\theta_{s1},\theta_{s2})$, with equal redshifts 
$z_{s1}=z_{s2}=z_{s3}=1$. Of course, we recover a behaviour similar to that
of Fig.~\ref{figr111_snap} obtained for the SNAP survey. Moreover, we
can check that the simple analytical model (\ref{stellar}) shows a good
agreement with the numerical simulations. We present in the four panels of
Fig.~\ref{figr111_red_sim} the dependence on the source redshift $z_{s3}$ of 
the correlator $r_{111}$, with two different redshifts $z_{s1}=1$ and 
$z_{s2}=2$. The dependence on $z_{s3}$ is very weak over the range 
$z_{s1}<z_{s3}<z_{s2}$ while $r_{111}$ decreases for larger $z_{s3}$, as 
expected. The agreement of the simple stellar model with numerical simulations 
is rather good for all cases.

\subsection{Constraints on the density field}
\label{Constraints on the density field}

We have seen in Fig.~\ref{figrpq_snap_20} that the correlation coefficients
$r_{pq}$ (or $r_{pqs}$) are almost independent of cosmological parameters.
Therefore, they can be used to monitor the noise of the survey or to constrain
the properties of the matter density field (e.g. the angular behaviour of the
many-body correlation functions). We have checked in 
Figs.~\ref{figrpq_sim}-\ref{figr111_red_sim} that the simple stellar model 
(\ref{stellar}) (which is a peculiar case of the more general class of 
hierarchical models) agrees rather well with numerical simulations. We now
show that this agreement is not due to the independence of the correlators
$r_{pqs}$ on the underlying density field. That is, although most of the 
dependence on cosmology has been factorized out of the ratios (\ref{rpq})
and (\ref{rpqs}) as well as the overall amplitude of the density fluctuations
(both properties are mostly probed by the one-point statistics), some
information on the detailed angular behaviour of the density correlations
remains in the correlation coefficients $r_{pqs}$. 

\begin{figure}
\protect\centerline{
\epsfysize = 2.7 truein 
\epsfbox[24 307 587 547]
{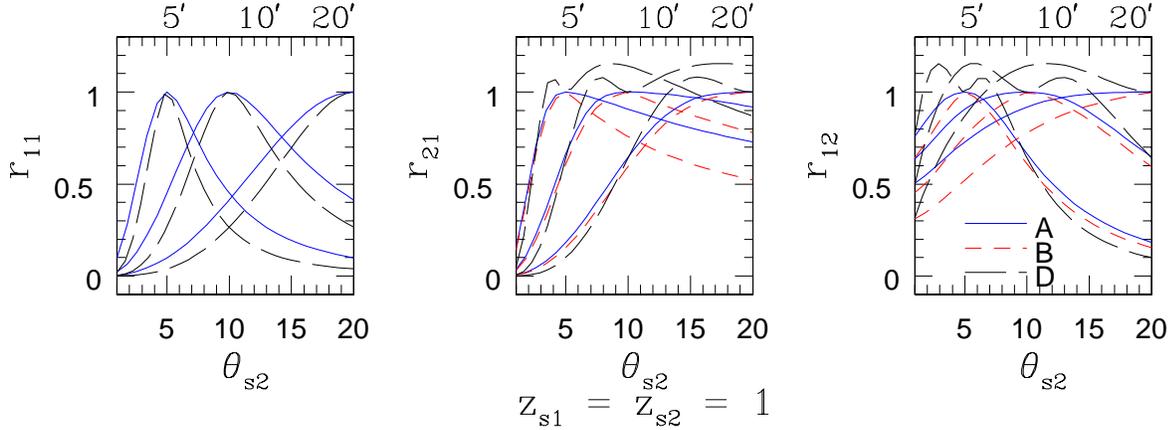}}
\caption{The two-point correlation coefficients $r_{pq}$ as in 
Fig.~\ref{figrpq_sim}, for $\theta_{s1}=5',10'$ and $20'$ and 
$z_{s1}=z_{s2}=1$. The different line-styles represent different analytical
models: the basic stellar model (A), the variant (B) (see main text) and the
Dirac-product model (D).}
\label{figrpq_app}
\end{figure}

\begin{figure}
\protect\centerline{
\epsfysize = 2.75 truein
\epsfbox[12 430 394 729]
{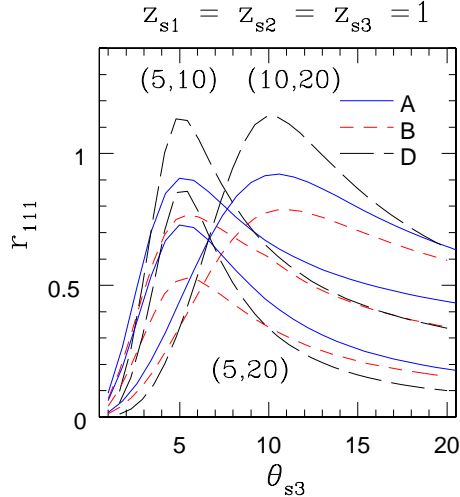}}
\caption{The three-point correlation coefficient $r_{111}$ as in 
Fig.~\ref{figr111_sim}, at $z_{s1}=z_{s2}=z_{s3}=1$ for three pairs of
smoothing angles $(\theta_{s1},\theta_{s2})$. The different line-styles 
represent different analytical models: the basic stellar model (A), the 
variant (B) (see main text) and the Dirac-product model (D).}
\label{figr111_app}
\end{figure}

To investigate this point we compare in 
Figs.~\ref{figrpq_app}-\ref{figr111_app} the predictions obtained from 
different analytical models. 
The solid line (A) is the stellar model (\ref{stellar}) which we
have used so far. Here we must note that this model is not fully defined
by eq.(\ref{stellar}) because the parameters $\tilde{S}_p$ vary with scale
and time. More precisely, the skewness parameter $\tilde{S}_3$ is obtained
by interpolating between the quasi-linear theory prediction and the HEPT
approximation (Scoccimarro et al. 1998), see Munshi et al. (2004). Then,
$\tilde{S}_3$ depends on the slope of the linear power-spectrum $P_L(k)$
and on the amplitude of the non-linear power $\Delta^2(k)$ at the wavenumber
$k=4/(\De\theta_s)$. However, for the two-point correlations we have two
angular scales $\theta_{s1},\theta_{s2}$ which can be different. Then, we 
extended the stellar model (\ref{stellar}) to these cases by using:
\beq
\mbox{model (A):} \;\;\; \tilde{S}_{p,q} = \tilde{S}_{p+q,0}^{p/(p+q)} 
\tilde{S}_{0,p+q}^{q/(p+q)} ,
\label{modelA}
\eeq
where we note $\tilde{S}_{p,q}$ the coefficient $\tilde{S}_{p+q}$ which appears
in a cross-product of the form $\lag X_1^p X_2^q \rag_c$, and the coefficient
$\tilde{S}_{p+p,0}$ (resp. $\tilde{S}_{0,p+q}$) is associated with only one 
angular scale $\theta_{s1}$ (resp. $\theta_{s2}$) where there is no ambiguity
for the typical wavenumber $k$. We have checked in 
sections~\ref{Two-point correlators},~\ref{Three-point correlators}
that this simple model (A) agrees quite well with numerical simulations.
Next, in order to evaluate the sensitivity of our predictions onto the
approximation (\ref{modelA}) we define the variant (B) by:
\beq
\mbox{model (B):} \;\;\; \tilde{S}_{p,q} = \min(\tilde{S}_{p+q,0}; 
\tilde{S}_{0,p+q}) .
\label{modelB}
\eeq
Thus the two models (A) and (B) coincide for $\theta_{s1}=\theta_{s2}$ and 
both can be described as stellar models of the form (\ref{stellar}).
The freedom between models (A) and (B), due to the scale dependence of
the parameters $\tilde{S}_p$, expresses the fact that the model (\ref{stellar})
is a simple phenomenological prescription but not a fully consistent model
(this would require a constant parameter $\tilde{S}_p$).
Finally, we also define a third model (D) where many-body density correlations
reduce to products of Dirac factors in real space:
\beq
\mbox{model (D):} \;\;\; \lag\delta(\bx_1) \dots\delta(\bx_p)\rag_c = 
\delta_D(\bx_2-\bx_1) \dots \delta_D(\bx_p-\bx_1) .
\label{modelD}
\eeq
We take the normalization coefficient in eq.(\ref{modelD}) to be constant
so that it cancels out of the correlation coefficients $r_{pq}$. For $p=2$
the model (D) merely corresponds to a white-noise power-spectrum, while
for all $p \geq 2$ it corresponds to the stellar model with a 
white-noise power-spectrum.
This model is not expected to be a good approximation of the matter density
field. Its only purpose is to check whether the detailed behaviour of
large-scale structures can be seen in the correlation coefficients $r_{pq}$.
It gives the contribution to the coefficients $r_{pq}$ provided by sheer
geometrical factors, since we have for identical source redshift distributions:
\beq
\mbox{model (D):} \;\;\; \lag X_1^p X_2^q \rag_c \propto 
\int\d{\vec \vartheta} \; U_1({\vec \vartheta})^p U_2({\vec \vartheta})^q 
\;\;\; \mbox{if} \;\;\; n_1(z_s)=n_2(z_s)  ,
\label{modelD2}
\eeq
where the normalization factor which only depends on $(p+q)$ cancels out
of the coefficients $r_{pq}$. The models (A), (B) and (D) extend in a
straightforward manner to three-point correlators like $r_{111}$.
Although the models (A), (B) and (D) yield different results despite their
similarity, we can note that the predictions obtained from model (A) for
the third-order correlators we study here are actually common to all models 
which obey
$\lag\delta(\bk_1)\delta(\bk_2)\delta(\bk_3)\rag_c \propto 
\delta_D(\bk_1+\bk_2+\bk_3) [ P(k_1) P(k_2) F(\bk_1,\bk_2) + \mbox{sym.} ]$
where $F(\bk_1,\bk_2)$ is any function of the angle between $\bk_1$ and 
$\bk_2$. Indeed, the angular dependence contained in $F(\bk_1,\bk_2)$ is
integrated out in the same manner for one-point, two-point and three-point
cumulants so that it cancels out of $r_{pqs}$. Of course, the behaviour
of $F(\bk_1,\bk_2)$ would show up if we consider different smoothing cells
which are not centered on the same line-of-sight. 

We show in Figs.~\ref{figrpq_app}-\ref{figr111_app} these three models
(A) (solid line), (B) (dashed line) and (D) (dot-dashed line). We can
see that the differences are quite apparent and are larger than the error-bars
obtained from numerical simulations in 
Figs.~\ref{figrpq_sim}-\ref{figr111_red_sim}. Therefore, although the 
correlation coefficients $r_{pq}$ and $r_{pqs}$ are almost insensitive to
cosmology they show a significant dependence on the properties of the
many-body density correlations (note that models (A) and (B) are quite
close). This separation between the background 
cosmology (which can be probed by one-point statistics) and the large-scale
structures themselves should be quite useful to discriminate between both
classes of phenomena and to probe gravitational clustering. In particular,
we can note that the differences between various models seen in 
Figs.~\ref{figrpq_app}-\ref{figr111_app} are sufficiently large to be
detected by future surveys like SNAP, see the error-bars in 
Figs.~\ref{figrpq_snap}-\ref{figr111_snap}. Finally, it appears that the
simple stellar model (A) defined from eqs.(\ref{stellar}),(\ref{modelA})
works best and provides a good description of the actual density
field for these purposes, as checked in 
sections ~\ref{Two-point correlators},~\ref{Three-point correlators}.

\section{Discussion}

We have extended the formalism developed in Munshi \& Coles (2002) to study 
the cross-correlations among various angular scales covering a range of 
redshift bins. We define the correlation coefficients $r_{pqs}$ in such a
way as to minimize the dependence on cosmological parameters and on the
redshift distribution of the sources. Since the aperture-mass window is 
very narrow in Fourier space they provide a useful tool to study gravitational 
clustering, through both the power spectrum and the bi-spectrum. In particular,
higher order coefficients $r_{pqs}$ where $p+q+s>2$ carry useful information 
about the level of non-Gaussianity in the projected density field. 
Generalizing our previous studies from two-point statistics to three-point 
statistics we have used $r_{111}$ as a diagnostics of the 
bi-spectrum. Indeed, although one-point or two-point cumulants or correlators
such as $r_{21}$ also contain useful information about the bi-spectrum, 
$r_{111}$ allows one to probe the bi-spectrum in more details. These tools
can be used not only to probe gravitational instability scenarios but
also to put constraints on primordial non-Gaussianities or higher-dimensional
theories of gravitation. We plan to study such issues in more detail in future
works. On the other hand, these estimators can also be used to monitor hidden 
systematics.

We have presented a detailed analysis of these correlation coefficients for
the wide SNAP survey, as an illustration for future weak-lensing surveys.
We have computed both the expected signal and its error-bars, taking into
account the intrinsic ellipticity of galaxies and the cosmic variance.
The cumulant estimators $H_{pqs}$ are built from the moment estimators 
$M_{pqs}$ so as to lower their scatter. Note that we take into account
the non-Gaussianities of the density field in the computation of the
error-bars. The scatter associated with SNAP class experiments is extremely 
small for second-order statistics such as the variance or the correlation
$r_{11}$ but it increases very fast with the order of the statistics. 
Nevertheless, these surveys are still able to probe third-order statistics 
with a reasonable accuracy. This should allow one to get meaningful constraints
on large-scale structures or on the observational noise.

Next, we have compared a simple stellar model with the results obtained from
numerical simulations. We had shown in previous papers that this simple 
analytical model agrees reasonably well with simulations for one-point
cumulants (e.g., Munshi et al. 2004). We have found in this work that it also
gives good predictions for two-point and three-point statistics over the
entire range of redshifts and smoothing scales of interest. Therefore,
this simple model provides a useful tool to investigate in great details
weak-lensing effects measured through smoothed statistics like the 
aperture-mass.

Finally, we have checked that the agreement between this simple model and
numerical simulations is not due to a weak sensitivity of these correlators
onto the structure of the underlying density field. Indeed, introducing
two other models for the detailed behaviour of the matter density correlations
we have found that the correlation coefficients $r_{pqs}$ show a significant
dependence on the properties of large-scale structures, which can be detected
in future surveys like SNAP. Therefore, such observations can yield 
constraints on the structure of the matter density field. Note however that
at order three the predictions obtained from the simple stellar model 
(\ref{stellar}) are shared by a wider class of models (e.g. all usual 
hierarchical models). In order to discriminate these various models one
would need to go up to fourth-order statistics (where one can draw several
hierarchical diagrams for instance) but error-bars become too large to give
good constraints.

In this paper we have only considered the cases where the smoothing 
patches are all lined up along the same radial direction, although they can 
have different redshift distributions or smoothing radii. It is trivial to 
generalise our results to cases where various patches are separated by a given 
separation angle. However with increasing separation the signal to noise will 
degrade which may prevent practical applications. We have only used the 
compensated filter associated with the aperture-mass which can directly be 
constructed from shear maps. Its presents the advantages of probing a narrow 
range of wavenumbers and of giving a non-zero signal for third-order
statistics. Nevertheless our formalism is completely general and we can 
replace this compensated filter by any window (like a top-hat for the 
convergence or a modified top-hat with angular dependence for shear 
components). However, in such cases the number of patches which can
carry completely independent information will decrease (because of the 
contribution of long wavelengths) which will increase observational error-bars.

\section*{acknowledgments}

DM was supported by PPARC of grant
RG28936. It is a pleasure for DM to acknowledge many fruitful
discussions with members of Cambridge Leverhulme Quantitative
Cosmology Group and members of Cambridge Planck Analysis Centre.
We would especially like to thank Andrew Barber for allowing
us to use the data from his simulations, generated for our previous 
collaborative works.

\appendix

\section{3pt Correlators}

As in Munshi \& Valageas (2004), the three-point cumulants
$\lag\Mapone^p\Maptwo^q\Mapthree^s\rag_c$ can be estimated from the estimators
$M_{pqs}$ defined by:
\beq
M_{pqs} = \frac{(\pi\theta_{s1}^2)^p (\pi\theta_{s2}^2)^q (\pi\theta_{s3}^2)^s}
{(N_1)_p (N_2)_q (N_3)_s} 
\sum_{\{(i_1,\dots,i_p);(j_1,\dots,j_p);(l_1,\dots,l_p)\}}^{\{N_1;N_2;N_3\}} 
Q_{i_1} \e_{{\rm t}i_1} \dots Q_{i_p} \e_{{\rm t}i_p} \;\; 
Q_{j_1} \e_{{\rm t}j_1} \dots Q_{j_q} \e_{{\rm t}j_q} \;\; 
Q_{l_1} \e_{{\rm t}l_1} \dots Q_{l_s} \e_{{\rm t}l_s} .
\label{Mpqs}
\eeq
Here, $N_1$, $N_2$ and $N_3$ are the number of galaxies within the patches 
$\pi\theta_{s1}^2$, $\pi\theta_{s2}^2$ and $\pi\theta_{s3}^2$ for the surveys 
$(1,2,3)$, while $N_{ij}$ and $N_{123}$ are the number of common galaxies to 
two or all three samples, and we noted 
$(N)_p = N (N-1) .. (N-p+1) = N!/(N-p)!$. In eq.(\ref{Mpqs}) we only sum over 
combinations of distinct galaxies and we assumed that $N_j \gg 1$. Thus, for
$M_{111}$ we sum over $N_1 N_2 N_3-N_{12}N_3-N_{13}N_2-N_{23}N_1+2N_{123}
\simeq N_1 N_2 N_3$ terms. We note 
$Q_j=Q_{\Map}({\vec \vartheta}_j)$ the value of the filter 
$Q_{\Map}({\vec \vartheta})$ at the location ${\vec \vartheta}_j$ of the 
galaxy $j$ and $\epsilon_{{\rm t}j}$ its tangential ellipticity. The filter 
$Q_{\Map}$ is given by:
\beq
Q_{\Map}({\vec \vartheta}) = \frac{\Theta(\vartheta<\theta_s)}{\pi\theta_s^2}
\; 6  \; \left(\frac{\vartheta}{\theta_s}\right)^2 
\left(1-\frac{\vartheta^2}{\theta_s^2}\right) .
\label{QMap}
\eeq
The mean and the dispersion $\sigma$ of $M_{pqs}$ are:
\beq
\lag M_{pqs}\rag = \lag\Mapone^p\Maptwo^q\Mapthree^s\rag , 
\;\;\; \sigma^2(M_{pqs})= \lag M_{pqs}^2\rag -\lag M_{pqs}\rag^2 .
\label{sigma}
\eeq
The scatter $\sigma(M_{pqs})$ only involves the variance 
$\sigma_*^2=\lag|\epsilon_*|^2\rag$ of the galaxy intrinsic ellipticities 
(even if the latter are not Gaussian). The latter enters the dispersion 
$\sigma^2$ through the combinations $\rho_j$ and $\rho_{ij}$ defined by:
\beq
\rho_j = \frac{5 N_j\lag M_{{\rm ap}j}^2\rag_c}{3\sigma_*^2} \;\;\; \mbox{and}
\;\;\; \rho_{ij} = \frac{5 N_i N_j \lag M_{{\rm ap}i} M_{{\rm ap}j}\rag_c}
{N_{ij} 3\sigma_*^2} .
\label{rho}
\eeq
The quantity $\rho_j$ is the signal-to-noise ratio of the weak-lensing 
variance over the random intrinsic ellipticity. It also measures the relative 
importance of the cosmic variance as compared with the galaxy intrinsic 
ellipticities to the scatter of relevant estimators (intrinsic ellipticities 
can be neglected if $\rho\gg 1$). 

The estimators $M_{pqs}$ defined in eq.(\ref{Mpqs}) correspond to a single
direction onto the sky, around which are centered the three smoothing windows 
of radii $\theta_{sj}$. In practice, we can average over $N_c$ such directions
on the sky with no overlap. This yields the estimators $\cM_{pqs}$ defined by:
\beq
\cM_{pqs} = \frac{1}{N_c} \sum_{n=1}^{N_c} M_{pqs}^{(n)} , \;\;\; \mbox{whence}
\;\;\; \lag \cM_{pqs} \rag = \lag M_{pqs} \rag = 
\lag\Mapone^p\Maptwo^q\Mapthree^s\rag \;\;\; \mbox{and} \;\;\; 
\sigma(\cM_{pqs}) = \frac{\sigma(M_{pqs})}{\sqrt{N_c}} ,
\label{cMpqs}
\eeq
where $M_{pqs}^{(n)}$ is the estimator $M_{pqs}$ for the cell $n$ and we 
assumed that these cells are sufficiently well separated so as to be 
uncorrelated. The estimators $M_{pqs}$ and $\cM_{pqs}$ provide a measure of 
the moments of weak lensing observables. However, as shown in 
Valageas et al. (2004b), it is better to first consider cumulant-inspired
estimators $H_{pqs}$ and $\cH_{pqs}$. For the case of the quantity 
$\lag\Mapone\Maptwo\Mapthree\rag_c$ the relevant estimator $H_{111}$ is:
\beq
H_{111} = M_{111} - \cM_{011} M_{100} - \cM_{101} M_{010} - \cM_{110} M_{001} ,
\;\;\; \cH_{111} = \frac{1}{N_c} \sum_{n=1}^{N_c} H_{111}^{(n)} , \;\;\; 
\mbox{whence} \;\;\; \lag \cH_{111} \rag = \lag\Mapone\Maptwo\Mapthree\rag_c .
\label{H111}
\eeq
Then the dispersion $\sigma^2(H_{111})$ reads:
\beqa
\lefteqn{\sigma^2(H_{111}) = \lag\Mapone^2\Maptwo^2\Mapthree^2\rag_c +
\lag\Maptwo^2\Mapthree^2\rag_c \lag\Mapone^2\rag_c 
\left[1+{1\over\rho_1}\right] + \lag\Mapone^2\Mapthree^2\rag_c 
\lag\Maptwo^2\rag_c \left[1+{1\over\rho_2}\right] } \nonumber \\ 
&& + \lag\Mapone^2\Maptwo^2\rag_c \lag\Mapthree^2\rag_c 
\left[1+{1\over\rho_3}\right] + \lag\Mapone^2\Maptwo\Mapthree\rag_c 
\lag\Maptwo\Mapthree\rag_c 2 \left[1+{1\over\rho_{23}}\right]
+ \lag\Mapone\Maptwo^2\Mapthree\rag_c \lag\Mapone\Mapthree\rag_c 2 
\left[1+{1\over\rho_{13}}\right] \nonumber \\ 
&& + \lag\Mapone\Maptwo\Mapthree^2\rag_c 
\lag\Mapone\Maptwo\rag_c 2 \left[1+{1\over\rho_{12}}\right]
+  2 \lag\Mapone\Maptwo^2\rag_c \lag\Mapone\Mapthree^2\rag_c +  2
\lag\Mapone^2\Maptwo\rag_c \lag\Maptwo\Mapthree^2\rag_c \nonumber \\ 
&& + 2 \lag\Mapone^2\Mapthree\rag_c \lag\Maptwo^2\Mapthree\rag_c + 3 
\lag\Mapone\Maptwo\Mapthree\rag_c^2 + \lag\Mapone^2\rag_c \lag\Maptwo^2\rag_c 
\lag\Mapthree^2\rag_c \left[1+{1\over\rho_1}\right] 
\left[1+{1\over\rho_2}\right] \left[1+{1\over\rho_3}\right] \nonumber \\ 
&& + \lag\Mapone^2\rag_c \lag\Maptwo\Mapthree\rag_c^2 
\left[1+{1\over\rho_1}\right] \left[1+{1\over\rho_{23}}\right]^2 + 
\lag\Maptwo^2\rag_c \lag\Mapone\Mapthree\rag_c^2 
\left[1+{1\over\rho_2}\right] \left[1+{1\over\rho_{13}}\right]^2 \nonumber \\ 
&& + \lag\Mapthree^2\rag_c \lag\Mapone\Maptwo\rag_c^2 
\left[1+{1\over\rho_3}\right] \left[1+{1\over\rho_{12}}\right]^2 +
\lag\Mapone\Maptwo\rag_c \lag\Mapone\Mapthree\rag_c \lag\Maptwo\Mapthree\rag_c
2 \left[1+{1\over\rho_{12}}\right] \left[1+{1\over\rho_{13}}\right]
\left[1+{1\over\rho_{23}}\right] 
\label{sigH111}
\eeqa
while the scatter of the estimator $M_{111}$ can be obtained from:
\beqa
\sigma^2(M_{111}) & = & \sigma^2(H_{111}) + 2 
\lag\Mapone^2\Maptwo\Mapthree\rag_c \lag\Maptwo\Mapthree\rag_c + 2 
\lag\Mapone\Maptwo^2\Mapthree\rag_c \lag\Mapone\Mapthree\rag_c \nonumber \\ 
&& + 2 \lag\Mapone\Maptwo\Mapthree^2\rag_c \lag\Mapone\Maptwo\rag_c + 
\lag\Mapone^2\rag_c \lag\Maptwo\Mapthree\rag_c^2 \left[1+{1\over\rho_1}\right] 
+ \lag\Maptwo^2\rag_c \lag\Mapone\Mapthree\rag_c^2 
\left[1+{1\over\rho_2}\right] \nonumber \\ 
&& + \lag\Mapthree^2\rag_c \lag\Mapone\Maptwo\rag_c^2 
\left[1+{1\over\rho_3}\right] + \lag\Mapone\Maptwo\rag_c 
\lag\Mapone\Mapthree\rag_c \lag\Maptwo\Mapthree\rag_c 
\left[6+{2\over\rho_{12}}+{2\over\rho_{13}}+{2\over\rho_{23}}\right] 
\label{sigM111} .
\eeqa
In order to obtain the dispersion of the correlation coefficient $r_{111}$
we also need the scatter of the cross-product 
$\sigma^2(H_{111};H_{300})=\lag H_{111} H_{300}\rag - \lag H_{111}\rag 
\lag H_{300}\rag$ which reads:
\beqa
\sigma^2(H_{111};H_{300}) & = & \lag\Mapone^4\Maptwo\Mapthree\rag_c + 
\lag\Mapone^3\Maptwo\rag_c \lag\Mapone\Mapthree\rag_c 3 
\left[1+{1\over\rho_{13}}\right] + \lag\Mapone^3\Mapthree\rag_c 
\lag\Mapone\Maptwo\rag_c 3 \left[1+{1\over\rho_{12}}\right] \nonumber \\ 
&& + \lag\Mapone^2\Maptwo\Mapthree\rag_c \lag\Mapone^2\rag_c 3 
\left[1+{1\over\rho_1}\right] + 3 \lag\Mapone^3\rag_c 
\lag\Mapone\Maptwo\Mapthree\rag_c + 6 \lag\Mapone^2\Maptwo\rag_c 
\lag\Mapone^2\Mapthree\rag_c \nonumber \\ 
&& + \lag\Mapone^2\rag_c \lag\Mapone\Maptwo\rag_c 
\lag\Mapone\Mapthree\rag_c 6 \left[1+{1\over\rho_1}\right] 
\left[1+{1\over\rho_{12}}\right] \left[1+{1\over\rho_{13}}\right] ,
\label{sigH111_H300}
\eeqa
while $\sigma^2(M_{111};M_{300})$ writes:
\beqa
\sigma^2(M_{111};M_{300}) & = & \sigma^2(H_{111};H_{300}) + 
\lag\Mapone^4\rag_c \lag\Maptwo\Mapthree\rag_c + \lag\Mapone^3\Maptwo\rag_c 
\lag\Mapone\Mapthree\rag_c + \lag\Mapone^3\Mapthree\rag_c 
\lag\Mapone\Maptwo\rag_c \nonumber \\ 
&& + 3 \lag\Mapone^2\Maptwo\Mapthree\rag_c 
\lag\Mapone^2\rag_c + \lag\Mapone^2\rag_c^2 \lag\Maptwo\Mapthree\rag_c
3 \left[1+{1\over\rho_1}\right] \nonumber \\ 
&& + \lag\Mapone^2\rag_c \lag\Mapone\Maptwo\rag_c 
\lag\Mapone\Mapthree\rag_c \left[6+{3\over\rho_{12}}+{3\over\rho_{13}}\right] .
\label{sigM111_M300}
\eeqa
We can check that as
for one-point and two-point cumulants, the estimators $H_{pqs}$ have a lower 
scatter than their counterparts $M_{pqs}$ and their dispersion only depends on
the combination $(1+1/\rho)$.

Finally, the three-point correlation coefficient $r_{111}$ can be estimated 
from the estimator $\cR_{111}$:
\beq
\cR_{111} = \frac{\cH_{111}}{\cH_{300}^{1/3} \cH_{030}^{1/3} \cH_{003}^{1/3}}
\;\;\;  \mbox{with} \;\;\; \lag\cR_{111}\rag \simeq r_{111}
\label{R111}
\eeq
and:
\beqa
\sigma^2(\cR_{111}) & \simeq & r_{111}^2 \Biggl \lbrace 
\frac{\sigma^2(\cH_{111})}{\lag\cH_{111}\rag^2} + \frac{1}{9} 
\frac{\sigma^2(\cH_{300})}{\lag\cH_{300}\rag^2} + \frac{1}{9} 
\frac{\sigma^2(\cH_{030})}{\lag\cH_{030}\rag^2} + \frac{1}{9} 
\frac{\sigma^2(\cH_{003})}{\lag\cH_{003}\rag^2} - \frac{2}{3} 
\frac{\sigma^2(\cH_{111};\cH_{300})}{\lag\cH_{111}\rag\lag\cH_{300}\rag} 
- \frac{2}{3} \frac{\sigma^2(\cH_{111};\cH_{030})}
{\lag\cH_{111}\rag\lag\cH_{030}\rag} \nonumber \\
&& - \frac{2}{3} \frac{\sigma^2(\cH_{111};\cH_{003})}
{\lag\cH_{111}\rag\lag\cH_{003}\rag} + \frac{2}{9} 
\frac{\sigma^2(\cH_{300};\cH_{030})}{\lag\cH_{300}\rag\lag\cH_{030}\rag}
+ \frac{2}{9} \frac{\sigma^2(\cH_{300};\cH_{003})}
{\lag\cH_{300}\rag\lag\cH_{003}\rag} + \frac{2}{9} 
\frac{\sigma^2(\cH_{030};\cH_{003})}{\lag\cH_{030}\rag\lag\cH_{003}\rag}
\Biggl \rbrace .
\label{sigR111}
\eeqa
Here we assumed that the scatter of $\cR_{111}$ is small and can be linearized.
The various terms in eq.(\ref{sigR111}) can be obtained from the expressions
derived above or from Munshi \& Valageas (2004).


\begin{thebibliography}{}

\bibitem{} Bacon D.J., Refregier A., Ellis R.S., 2000, MNRAS, 318, 625
\bibitem{} Barber A. J., Munshi D., Valageas P., 2004, MNRAS, 347, 665
\bibitem{} Bernardeau F., Schaeffer R., 1992, A\&A, 255, 1
\bibitem{} Bernardeau F., Valageas P., 2000, A\&A, 364, 1
\bibitem{CBT} Couchman H. M. P., Barber A. J., Thomas P. A., 1999, MNRAS, 308, 180
\bibitem{} Bernardeau F., Valageas P., 2000, A\&A, 364, 1
\bibitem{} Bernardeau F., Colombi S., Gaztanaga E., Scoccimarro R., Phys.Rept., (2002), 367, 1
\bibitem{CTP} Couchman H. M. P., Thomas P. A., Pearce F. R., 1995, ApJ., 452, 
\bibitem{} Fry J.N., 1984, ApJ, 279, 499
\bibitem{} Hoekstra H., Yee H. K. C., Gladders M. D., 2002, ApJ, 577, 595
\bibitem{} Jain B., Seljak U., 1997, ApJ, 484, 560
\bibitem{} Jain B., Seljak U., White S.D.M., 2000, ApJ, 530, 547
\bibitem{} Jaroszyn'ski M., Park C., Paczynski B., Gott J.R., 1990, ApJ, 365, 22
\bibitem{} Kaiser N., 1998, ApJ, 498, 26
\bibitem{} Munshi D., Coles P., Melott A.L., 1999a, MNRAS, 307, 387
\bibitem{} Munshi D., Coles P., Melott A.L., 1999b, MNRAS, 310, 892
\bibitem{} Munshi D., Melott A.L., Coles P., 1999, MNRAS, 311, 149
\bibitem{} Munshi D., Coles P., 2003, MNRAS, 338, 846
\bibitem{} Munshi D., Jain B., 2000, MNRAS, 318, 109
\bibitem{} Munshi D., Jain B., 2001, MNRAS, 322, 107
\bibitem{} Munshi D., Valageas P. \& Barber A., 2004, MNRAS, in press
\bibitem{} Munshi D., Valageas P. 2004, MNRAS, in press
\bibitem{} Peacock J.A., Dodds S.J., 1996, MNRAS, 280, L19
\bibitem{} Peacock J. A., Smith R. E., 2000, MNRAS, 318, 1144
\bibitem{} Schneider P., Weiss A., 1988, ApJ, 330,1
\bibitem{} Schneider P., Van Waerbeke L., Jain B., Kruse G., 1998,
\bibitem{} Stebbins A., 1996, astro-ph/9609149
\bibitem{} Szapudi I., Szalay A.S., 1993, ApJ, 408, 43
\bibitem{} Szapudi I., Szalay A.S., 1997, ApJ, 481, L1
\bibitem{} Takada M., Jain B., 2003, MNRAS, 344, 857
\bibitem{} Valageas P., 2000a, A\&A, 354, 767
\bibitem{} Valageas P., 2000b, A\&A, 356, 771
\bibitem{} Valageas P., Barber A. J., Munshi D., 2003, MNRAS
(in press) (astro-ph/0303472)
\bibitem{} Van Waerbeke L., Bernardeau F., Mellier Y., 1999,
A\& A, 342, 15
\bibitem{} Van Waerbeke L., Mellier Y., Pell\'o R., Pen U.-L., McCracken 
H. J., Jain B., 2002, A\&A, 393, 369
\bibitem{} Villumsen J.V., 1996, MNRAS, 281, 369
\bibitem{} Wambsganss J., Cen R., Ostriker J.P., 1998, ApJ, 494,
298
\bibitem{} Wambsganss J., Cen R., Xu G., Ostriker J.P., 1997, ApJ, 494,
29

\end{thebibliography}
\end{document}